\documentstyle[preprint,revtex]{aps}
\begin{document}
\draft
\centerline{\hfill CERN-TH.6658/92}
\centerline{\hfill NUHEP--TH--92--17}
\pagestyle{empty}
\begin{title}
CP violation in top pair production\\
at an $e^+e^-$ collider
\end{title}
\author{
Darwin Chang$^{(1,2)}$,
Wai--Yee Keung$^{(3,4)}$\\
Ivan Phillips$^{(1)}$\\
\quad\\  \quad \\
}
\begin{instit}
$^{(1)}$Department of Physics and Astronomy,\\
Northwestern University, Evanston, IL 60208, USA.\\
$^{(2)}$Institute of Physics, Academia Sinica, Taipei, Taiwan, R.O.C.\\
$^{(3)}$Physics Department, University of Illinois at Chicago,
IL 60680, USA \\
$^{(4)}$Theory Division, CERN, CH-1211, Geneva 23, Switzerland \\
\end{instit}
\vskip 1cm
\centerline{\bf Abstract}
\vskip -.5cm
\begin{abstract}
We investigate a possible CP violating effect in $e^+e^-$
annihilation into $t\bar t$ top quark pairs.
As an illustrative example, we assume the source of the
CP nonconservation is in the Yukawa couplings of a neutral Higgs boson
which contain both scalar and pseudoscalar pieces.
One of the interesting observable effects is the difference in production
rates between the two CP conjugate polarized $t\bar t$ states.
\end{abstract}
\vskip -1cm
\centerline{\hskip -.3in CERN--TH.6658/92  \hfill}
\centerline{\hskip -.3in September 1992    \hfill}
\newpage
%
%
\narrowtext
\setcounter{page}{1}
\pagestyle{plain}

Since the top quark is widely believed to be within the reach of the present
collider machines, it is not unreasonable for theorists to imagine what we
can learn from the top quark.  The best place to study the top quark in detail
is in an $e^+e^-$ collider.  One of the facts one would like to learn from the
discovery of the top quark is the origin of the still mysterious CP violation.
In this paper we investigate a way CP violation can manifest itself in the
top pair production of an $e^+e^-$ collider.

Among the various mechanisms of CP violation the one that may manifest
itself most easily is the neutral Higgs mediated CP violation.  Since
the neutral Higgs couplings are typically proportional to the quark
mass, the large mass of the top quark naturally gives large couplings to
the neutral Higgs bosons.  In addition, the top quark, due to its short
lifetime, is widely believed to decay before it
hadronizes. Therefore the information about its
polarization may be preserved in its decay products.  If that is the
case, then one can investigate the source of CP nonconservation by
measuring the CP violating observable involving a polarized top pair in
the final state.  This idea of detecting the rate asymmetry between
different polarized states was recently pursued in
Refs.~\cite{ref:Peskin,ref:DK}.
For $t {\bar t}$ production through the virtual photonic or Z
intermediate states, to the lowest
order in the final state quark mass, the polarizations of the quarks are
either $t_L {\bar t}_R$ or $t_R {\bar t}_L$. (Note that we have adopted
the notation that ${\bar t}_L$ is the antiparticle of $t_R$ and should
be left handed.) These two modes are CP self-conjugate.  However since the
top quark is heavy, there will also be large percentage of
$t_L {\bar t}_L$ and $t_R {\bar t}_R$ modes which are
CP conjugates of each
other. Therefore one can consider the CP asymmetry in the
event rate difference, $N(t_L {\bar t}_L)-N(t_R {\bar t}_R)$.

For detection of the asymmetry $N(t_L {\bar t}_L)-N(t_R {\bar
t}_R)$\cite{ref:Peskin,ref:DK}, one assumes that
the $t$ quark decays semileptonically through the usual $V-A$ weak
interaction.
Assuming that the
hadronization time is much longer than the decay time\cite{ref:Bigi},
one can analyze polarization dependence of its decay at the quark level.
The top quark first decays into a $b$ quark and a $W^+$ boson,
which subsequently becomes $\l^+ \nu$.
For heavy top quark, the $W^+$ boson produced in top decay is predominantly
longitudinal.
Due to the $V-A$ interaction, the $b$ quark is preferentially produced
with left-handed helicity.  So the longitudinal $W^+$ boson is
preferentially produced along the direction of the top quark polarization.
Therefore the anti--lepton $\l^+$ produced in the $W^+$ decay is also
preferentially in that direction.
In the rest
frame of the $t$, the angular distribution\cite{ref:Kuhn}
of the produced $\l^+$ has the
form $1+\cos\psi$, with $\psi$ as the angle between $\l^+$ and the
helicity axis of the $t$.
Above the $t {\bar t}$ threshold, the top quark is produced with nonzero
momentum. As a result of the Lorentz boost, the anti--lepton
$\l^+$ produced in the decay of the right handed top quark $t_R$ has
a higher energy than that produced in the decay of the left handed top
quark $t_L$. Similarly, the lepton $l$ produced in the decay of ${\bar
t}_L$ has a higher energy than that produced in the decay of  ${\bar
t}_R$. Consequently, in the decay of the pair $t_L {\bar t}_L$ the
lepton from $\bar t_L$ has a higher energy than the anti--lepton
from $t_L$; while in the decay of $t_R {\bar t}_R$ the anti--lepton
has a higher energy.
Therefore one can observe $N(t_L {\bar t}_L)-N(t_R {\bar t}_R)$
by measuring the energy asymmetry in the resulting
leptons\cite{ref:OtherCP}.

In order to generate the asymmetries
$N(t_L {\bar t}_L)-N(t_R {\bar t}_R)$
it is necessary to include effects of the final state interactions
in order to escape from the hermiticity constraint at the tree--level
due to the CPT theorem.  In our case both CP violation and the final state
effect are produced by the same one--loop graphs.

CP non--conservation occurs in the complex Yukawa coupling,
\begin{equation} {\cal L}_{CPX}=-(m_t/v) \bar t (AP_L+A^*P_R) t H
   +  B H (m_Z^2/v) Z^\nu Z_\nu
\;.
\end{equation}
Here $v=(\sqrt{2} G_F)^{-{1\over 2}} \simeq 246 \hbox{ GeV}$.
The complex coefficient $A$ is a combination of model-dependent mixing angles.
Simultaneous presence of both the real part $A_R=\hbox{Re} A$ and the
imaginary part $A_I=\hbox{Im} A$ guarantees CP asymmetry. For example, in
the low energy regime, it can give rise to the electric dipole moment of
elementary particles \cite{ref:Weinberg,ref:Barr}. Here we will show
that CP nonconservation manifests itself in the event rate difference in
collider experiments.

The vertex amplitude $ ie \Gamma^j$ for the virtual
$\gamma^*$ or $Z^*$ turning into $t(p)$ and $\bar t(p')$
is parametrized in the following expression:
\begin{equation}
  \Gamma^j_\mu  = c_v^j \gamma_\mu  + c_a^j \gamma_\mu\gamma_5
                + c_d^j i\gamma_5 {p_\mu-p'_\mu\over 2 m_t} + \cdots,
\quad j=\gamma,Z.
\label{eqn:form}
\end{equation}
We use the tree--level values for  $c_v$ and $c_a$. They are
\begin{eqnarray}
  c_v^\gamma=&\case2/3, \quad\quad c_a^\gamma=0,          \nonumber\\
   c_v^Z=& (\case 1/4 - \case 2/3 x_W)/\sqrt{x_W(1-x_W)}\;,       \\
   c_a^Z=& -\case 1/4                 /\sqrt{x_W(1-x_W)}
\; .  \nonumber
\end{eqnarray}
The $c_d$ term is the the electric dipole moment factor. The spinor
structure can be recast into another familiar form
$\sigma_{\mu\nu}(p+p')^\nu\gamma_5/2m_t$.
It is induced at the one--loop level as shown in the Fig.~1. We are
interested in the absorptive parts which can be easily calculated
according to the Cutkosky rules.  It can also be easily shown that,
in the limit that the electron mass is ignored,
Im$c_d^j$ are the only relevant one loop form factors for the CP
violating quantities we are interested in.

The leading contribution to Im$c_d^\gamma$ comes from the rescattering
of the top quark pair through the Higgs--boson exchange.
\begin{equation}
\hbox{Im} c_d^\gamma = c_v^\gamma ({m_t\over v})^2
                           {A_RA_I t^2\over 2\pi\beta}
                       (1-{h^2\over \beta^2} \log(1+{\beta^2\over h^2}))
\;.
\end{equation}
The dimensionless variables are defined by,
$t=m_t/\sqrt{s}$, $z=m_Z/\sqrt{s}$, and $h=m_H/\sqrt{s}$.
For Im$c_d^Z$, there is a similar contribution. In addition,
there could be a contribution due to the $ZH$ intermediate state, Fig~1b,
provided the kinematics is allowed.
\begin{equation}
\hbox{Im} c_d^Z= {c_v^Z \over c_v^\gamma}\hbox{Im} c_d^\gamma
            -{ \alpha A_IBc_v^Z t^2 \over 2(1-x_W)x_W\beta^3 }
           [\beta \beta_Z+(2t^2+2t^2h^2-2t^2z^2-h^2)L]
\;.
\label{eq:zedm}
\end{equation}
Here $\beta_Z^2=1+h^4+z^4-2z^2-2h^2-2h^2z^2$, $\beta^2=1-4t^2$, and
the logarithmic factor
$L=\log(L_-/L_+)$ with $L_\pm=1-z^2-h^2 \pm \beta\beta_Z$.
Other irrelevant terms, like the magnetic moments, are not listed in
Eq.(\ref{eqn:form}). Note that there is no CP violating contribution
due to $c_a^Z$ coupling in the limit that the electron mass is ignored.
Our expression in Eq.(\ref{eq:zedm}) agrees with
that in Ref.\cite{ref:Bernreuther,ref:BMS}.
The amplitudes for the process $e^+e^-\rightarrow t\bar t$
of different helicities have been given in the literature\cite{ref:KLY}.
Now, we can obtain the explicit CP asymmetry in the difference of the
production rates,
$$
  {\ \hskip -1.5cm}
  \delta \equiv [N(t_L\bar t_L) -N(t_R\bar t_R)]/N(t\bar t; \hbox{all})
$$
\begin{equation}
     \hskip 1.5in
= { \sum_{i=L,R} 2\beta[(1-z^2)c_v^\gamma + r_i c_v^Z]
  [(1-z^2)\hbox{Im}c_d^\gamma + r_i \hbox{Im}c_d^Z]  \over
\sum_{i=L,R} (3-\beta^2)[(1-z^2)c_v^\gamma + r_i c_v^Z]^2 +
               2\beta^2 r_i^2  {c_a^Z}^2}
\;.
\label{eq:asym}
\end{equation}
Here $r_i=({1\over 2}\delta_{i,L}-x_W)/\sqrt{x_W (1-x_W)}$, which
is the $Z$--coupling of the electron of different chiralities.
Typical values of $\delta$ are shown in Fig.~2.

We can make use of this asymmetry parameter $\delta$ to illustrate the
the difference in the energy distributions of $l^+$ or $l^-$
from the $t$ or $\bar t$ decays.
The energy $E_0(l^+)$ distribution of a static $t$ quark decay
$t\rightarrow l^+\nu b$ is very simple\cite{ref:Kuhn}
in the narrow width $\Gamma_W$
approximation when $m_b$ is negligible.
\begin{equation}
    f(x_0)= \left\{
              \begin{array}{lll}
       x_0 (1-x_0)/D & \quad\  & \mbox{if $m_W^2/m_t^2 < x < 1$}, \\
       0             & \quad\  & \mbox{otherwise.}
             \end{array}
              \right.
\end{equation}
Here we denote the scaling variable $x_0=2E_0(l^+)/m_t$ and
the normalization factor $D={1\over 6}-{1\over 2}(m_W/m_t)^4
+{1\over 3}(m_W/m_t)^6$.
When the $t$ quark is not static, but moves at a speed $\beta$
with helicity $L$ or $R$, the distribution expression becomes
a convolution,
\begin{equation}
     f_{R,L}(x,\beta)=
    \int_{x/(1+\beta)}^{x/(1-\beta)} f(x_0)
            {\beta x_0 \pm (x-x_0) \over 2 x_0^2\beta^2}
      dx_0
\;.
\end{equation}
Here $x=2E(l^+)/E_t$.
Similar distributions for the $\bar t$ decay is related by CP
conjugation at the tree--level.
Using the polarization asymmetry formula in  Eq.(\ref{eq:asym}),
we can derive an expression for the difference in the energy
distributions of $l^-$ and $l^+$:
\begin{equation}
{1\over N} \Bigl[ {dN\over dx(l^+)}-{dN\over dx(l^-)} \Bigr]
=\delta[f_L(x,\beta)-f_R(x,\beta)]
\;.
\label{eq:disd}
\end{equation}
Here distributions are compared at the same energy for the lepton
and the anti--lepton, $x(l^-)=x(l^+)=x=4E(l^\pm)/\sqrt{s}$.
The count $N$ includes events with
prompt leptons or anti--leptons from the top pair production.
It is useful to compare
Eq.(\ref{eq:disd}) with that of the overall energy distribution,
$$
{1\over N} \Bigl[ {dN\over dx(l^-)}+{dN\over dx(l^+)} \Bigr]
={  \sum_{i=L,R} 4\beta r_i c_a^Z[(1-z^2)c_v^\gamma +r_i c_v^Z]
                                       [f_R(x,\beta)-f_L(x,\beta)]
                        \over
\sum_{i=L,R} (3-\beta^2)[(1-z^2)c_v^\gamma + r_i c_v^Z]^2 +
               2\beta^2 r_i^2  {c_a^Z}^2}
$$
\begin{equation}
 \quad\quad
           +  f_L(x,\beta) + f_R(x,\beta)
\;.
\label{eq:diss}
\end{equation}
Here we only keep the dominant tree--level contribution.
Fig.~3 shows the overall prompt lepton energy distribution
of Eq.(\ref{eq:diss}),
and the ratio of the expressions in Eq.(\ref{eq:disd}) and
Eq.(\ref{eq:diss}).

In conclusion, we have shown that the CP  violation in top pair
production in $e^+e^-$ annihilation can be at the level of
$10^{-3}$. Experiments can measure the asymmetry in the energy
distributions of prompt leptons and anti--leptons. An $e^+e^-$ machine
of high luminosity is needed to accumulate several million
prompt lepton events from the $t\bar t$
production in order to search for the CP nonconservation.

D.C. wishes to thank the Theory Group at the
Institute of Physics at Academia Sinica in Taipei, Taiwan and
Department of Physics and Astronomy at University of Hawaii at Manoa
for hospitality
while this work was in progress.  This work is supported by grants from
Department of Energy and from National Science Council of Republic of China.

\figure{Feynman diagrams for the  process $e^+e^-\rightarrow t\bar t$.
The tree amplitude interferes with those one--loop amplitudes with
(a) the final state interactions due to the exchange of a Higgs boson, or
(b) the intermediate state production of the $ZH$ bosons.
}
\figure{
$[N(t_L \bar t_L) - N(t_R \bar t_R)] / N(\hbox{all } t\bar t)$
versus $\sqrt{s}$  in the solid (dotted, dashed) curve for the case
that $m_H= 50$ (150,200) GeV and $m_t=100$ GeV.
The parameters are chosen to be $A_I=A_R=B=1$.
}
\figure{The energy distribtuions of prompt leptons, for
the case that $m_t=120$ GeV, $m_H=50$ GeV, $\sqrt{s}=300$ GeV, and
$A_I=A_R=B=1$. Case (a) for $N^{-1}[dN/dx(l^+)+dN/dx(l^-)]$,
and case (b) for $[dN/dx(l^+)-dN/dx(l^-)]/[dN/dx(l^+)+dN/dx(l^-)]$.
}
\end{document}